\documentclass[preprint,prl,showpacs,floatfix]{revtex4}

\usepackage{graphicx}

\begin{document}

\title
{
%Shear-Thickening and Thinning-like behaviors in a Simple Lattice Gas
%System
Complex Transport Phenomena in a Simple Lattice Gas System
}
    
\author
{ 
Akinori Awazu\footnote{E-mail address: awa@daisy.phys.s.u-tokyo.ac.jp}
}

\affiliation
{
Department of Physics, University of Tokyo, Hongou 7-3-1, Bunkyou-ku, Tokyo 113-0033, Japan.
}

\date{\today}

\begin{abstract}
The transport phenomena of a nonequilibrium lattice gas system are
investigated. We consider a simple system that consists of two particles
interacting repulsively and the potential forces acting on these
particles. Under an external driving field applied to only one
particle, we found the following relation between the mean velocity 
of the driven particle and the coefficient of effective drag of this 
particle under certain conditions; With the increase in the mean 
velocity, the coefficient of effective drag varies in the form, 
increase $\to$ decrease $\to$ increase $\to$ decrease. Moreover, under 
other conditions, we found the following relations between these values 
which show changes similar to those between shear rate and shear 
viscosity observed in the shear-thickening polymer or colloidal 
suspensions; With the increase in the mean velocity, the coefficient 
of effective drag varies in the form, increase $\to$ decrease or 
decrease $\to$ increase $\to$ decrease. We explain the mechanisms of 
such phenomena by considering the transition diagrams.  
\end{abstract}


\maketitle

\section{introduction}
Nonequilibrium lattice gases are simple mathematical models, which have
been useful and important in studies of the several properties of 
nonequilibrium systems with numerous degrees of freedom\cite{ss1}. 
Recently, many varieties of nonequilibrium phenomena such as
nonequilibrium phase transitions\cite{ss2}, a variety of particle 
flows\cite{ASEP,ASEP1,2lane0,2lane1}, appearances of long-range 
spatial correlations\cite{ss3,hal}, and the mathematical foundations of
nonequilibrium statistical mechanics and
thermodynamics\cite{ss4,ss5,ss55,ss6} have been investigated through
such lattice gases.

In this paper, we investigate the transport behaviors of a lattice gas
system with a periodic boundary, which consists of only two particles
interacting repulsively and the potential forces acting on them. Under
nonequilibrium conditions, lattice gas systems have been known to show
some nontrivial phenomena, such as the appearance of long-range spatial
correlations\cite{hal} and anomalous drift motions\cite{2lane1}, even if
the system involves only two particles. 

In our system, the following complex transport properties are found when
only one particle is driven by an external driving field; With the
increase in the mean velocity of the driven particle, the coefficient of
effective drag of this particle (=[driving field strength]/[mean
velocity]) varies in the form, increase $\to$ decrease $\to$ increase
$\to$ decrease under certain conditions. Moreover, under other
conditions, the relation between these values show changes similar to
those between shear rate and shear viscosity observed experimentally 
in the shear-thickening polymer\cite{poly0,poly1,poly11} or colloidal
suspensions\cite{col0,col1}; With the increase in the mean velocity, the
coefficient of effective drag varies in the form, increase $\to$
decrease or decrease $\to$ increase $\to$ decrease.

In the following sections, details of the numerical and analytical
results are shown. First, we introduce our model and show the numerical
results, where the above-mentioned phenomena are observed. Next, we
explain the mechanism of these phenomena by considering the transition
diagrams. Last, we show the similarities between our obtained results
and those observed in the polymer solutions or colloidal suspensions. 

\section{Model}
Now, we introduce a lattice gas model, which is the same as
that studied in our previous paper\cite{2lane1}. We consider a lattice
system with two parallel one-dimensional lanes where each lane involves
$L$ sites with a periodic boundary. Each lane contains only one particle
which moves randomly to the nearest sites without changing lanes. The
sites occupied by particles in the $1$st and $2$nd lanes are denoted
$x_1$ and $x_2$, respectively, which are given as integer numbers from
$0$ to $L-1$.

The effect of potential forces acting on the particles is described
by the following Hamiltonian:
\begin{equation}
H(x_1,x_2)=V(x_1)+V(x_2)+V_{12}(x_1, x_2),
\end{equation}
where $V(x)$ represents the one-body potential on each lane, and
$V_{12}(x_1, x_2)$ represents the interaction potential between
the two particles. Furthermore, an external driving field is applied to
the particle on the $2$nd lane. We denote the field strength $F$.

The time evolution of this system is described by the iteration of
the following three steps. First, one of the two particles is randomly 
chosen. Let the position of the chosen particle be $x$. Second,
its neighboring site $y$, $x-1$ or $x+1$, is randomly chosen.
Third, the chosen particle moves from $x$ to $y$ with the following 
probability
\begin{equation}
c(x,y;x_1,x_2)=\frac{1}{1 + \exp[Q(x \to y; x_1, x_2)/k_{B}T]},
\end{equation}
with
\begin{equation}
Q(x \to y ; x_1, x_2) = H(x_1', x_2') - H(x_1, x_2) - F(x_2'-x_2),
\end{equation}
where $(x_1',x_2')=(x_1,y)$ when $x=x_2$, and $(x_1',x_2')=(y,x_2)$ when
$x=x_1$\cite{periodic}. 
$T$ is temperature and the Boltzmann constant $k_{B}$ is set 1.
Here, the time step is given by [No. of above iterations]/[No. of
particles ($=2$)].
%\cite{metro}.

Specifically, we study the case where $V(x)=V |L/2-x|$ (Fig. 1), and 
$V_{12}(x_1,x_2)= I \delta_{x_1,x_2}$ using the $L \times L$ unit matrix
$\delta_{ij}$. 
Also, we focus on the case $L=4$. We found that this size is the minimum
required to exhibit the phenomenon we demonstrate in the presented
paper. 

\section{Simulation results}
In this section, we show the simulation results of this system. 
We mainly focus on the cases with $V < I$ and $|F| < I+V$ because 
we obtained some novel phenomena as shown in the belows. Such conditions
indicate that the influences of the potential forces and the driving
field are weak compared to those of the particle-particle interactions.
In order to characterize the system, we define the
mean velocity of the driven particle (in the $2$nd lane) in steady state,
$u$, as the difference of the long time average of the moving ratio of
the driven particle in the positive and negative directions. Here, the
direction $x_{i}: 0 \to 1 \to ... \to (L-1) \to 0 \to$ is positive. For
simplicity, $I = 1$ and $F > 0$ are set.

First, we focus on the cases where $T$ is small enough compared to $I$. 
Figures 2(a), (b) and (c) show $u$ as a function of $F$ for 
(a) $V = 0.25$ with $T = 0.05$ or $T = 0.07$, (b) $V = 0.4$ with 
$T = 0.05$ or $T = 0.08$, and (c) $V = 0.6$ with 
$T = 0.08$ or $T = 0.1$. Then, the following two types of $F - u$
profiles appear; A) $u$ increases slowly $\to$ steeply $\to$
slowly $\to$ steeply with the increase in $F$ as shown in (a) and (b)
with a smaller $T$. B) $u$ increases steeply $\to$ slowly $\to$
steeply with the increase in $F$ as shown in (b) with a larger $T$ and (c).

From these results, the relations between $u$ and the coefficient of
effective drag of the driven particle, $\eta$, defined as $F / u$ are
straightforwardly obtained. 
Figures 3(a), (b) and (c) show $\eta$ as a function of $u$ for 
(a) $V = 0.25$ or $V = 0.3$, (b) $V = 0.35$ or $V = 0.4$, and 
(c) $V = 0.55$ or $V = 0.6$. Here, the properties of $u - \eta$ profile
change depending on $V$ and $T$ as following. 

I) -i) In the case with $V < I/3$, as shown in Fig. 3(a), $\eta$ varies in the
form, increase $\to$ decrease $\to$ increase $\to$ decrease, with the
increase in $u$ for small $T$ (for example $T = 0.05$). -ii) On the other 
hand for a little larger $T$ (for example $T = 0.07$), the change in 
$\eta$ becomes less sharp and changes in the form, increase $\to$ fast
decrease $\to$ slow decrease $\to$ fast decrease, with the increase in
$u$. Here, $u$ at the maximum values of $\eta$ decrease with the
increase in $V$ or the decrease in $T$.

II) -i) In the case with $I/3 < V < I/2$, as shown in Fig. 3(b), $\eta$
varies in the form, increase $\to$ decrease $\to$ increase $\to$
decrease, with the increase in $u$ for small $T$ (for example $T = 0.05$). 
-ii) On the other hand for a little larger $T$ (for example $T = 0.08$), 
the change in $\eta$ becomes less sharp, and simpler in the form, 
decrease $\to$ increase $\to$ decrease, with the increase in 
$u$. Here, $u$ at the maximum values of $\eta$ decrease with 
the increase in $V$ or the decrease in $T$. 
   
III) In the case with $I/2 < V$, as shown in Fig. 3(c), $\eta$ varies in the
form, decrease $\to$ increase $\to$ decrease, with the increase in $u$
independently of $T$. Here, $u$ at the maximum values of $\eta$ decrease
with the decrease in $V$ or $T$. 

It is remarkable that the $u - \eta$ profile given in case I-ii) 
appears qualitatively similar to that between 
the shear rate and shear viscosity coefficient of the shear-thickening 
polymer solutions obtained experimentally\cite{poly0,poly1,poly11}.
On the other hand, in the case II-ii) or III), 
the $u - \eta$ profile appear qualitatively similar to
that between the shear rate and shear viscosity coefficient of the 
shear-thickening colloidal suspension obtained
experimentally\cite{col0,col1}. 

Phase diagram for the properties of $u - \eta$ profile as functions of
$V$ and $T$ is given as Fig. 3(d). Here, the boundaries given as $V = I/3$
and $V < I/2$ (as indicated by the broken lines.) are predicted by the 
analytical studies as mentioned in the next section. If $T$ is given
larger, $u - \eta$ profiles for several $V$ become less sharp and close
to flat like Newtonian fluid continuously as shown in Fig. 3(a), (b) and
(c) (for example $T = 0.2$ or $T = 0.32$). 

It is noted that, if $V >> I$ in which the particle-particle interactions
are much smaller than any other effects, $u$ is roughly estimated as 
$\propto \exp(-(2V-I-F)/T)- \exp(-(2V-I+F)/T)$. In such cases, $\eta$ 
decreases monotonically with the increase in $u$ for small $T$, and
$\eta - u$ profile closes to flat with the increase in $T$. Thus, the
particle-particle interactions play important roles to show the
presented non-trivial behaviors like those observed shear-thickening
solutions.

\section{Analysis of the model by transition diagrams}
In this section, we try to explain the mechanism of our obtained
phenomena by considering the transition diagrams.. First, we name all
states of this system using the sites occupied by the particles in each
lane $(x_{1}, x_{2})$. Since $L=4$, this system has 16 states from 
$(0, 0)$ to $(3, 3)$. The transitional tendency from $(x_{1}, x_{2})$ to
$(x_1',x_2')$ is given by $Q(x \to y ; x_1, x_2)$, which was defined
previously. Here, $(x_1',x_2') = (x_1,y)$ when $x = x_2$, and
$(x_1',x_2') = (y,x_2)$ when $x=x_1$. Then, we obtain the transition
diagrams; including all information on the transitions between states. 

We focus on the $F$ dependent changes of the transition diagrams. 
Figure 4 shows the typical transition diagrams in the cases where 
(a) $F < V$, (b) $V < F < I - 2V$, (c) $I - 2V < F < I - V$, 
(d) $I - V < F < I$ and (e) $I < F$ under $V < I/3$. Here, $(i, j)$
indicates the state $(x_{1}, x_{2})$, the arrows give the direction of
the transition with a probability higher than $\sim 0.5$ implying 
$Q(x \to y ; x_1, x_2) \stackrel{<}{\sim} 0$, and the values beside
these arrows indicate the value of $|Q(x \to y ; x_1, x_2)|$.  
The motion of the system is represented by the random walk caused by
the fluctuation with $T$ in the transition diagram. Here, 
the transition in a vertical direction indicates the 
motion of the particle in the $1$st lane, and that in a horizontal 
direction indicates the motion of the driven particle in the $2$nd lane. 

As shown in Figs. 4, there exist two types of states, the transient
state, which has some arrows directing to some other states, and the
attractor, which has no arrows directing to any other states. If $T$ is
small enough compared to $I$ and $V$, the transitions from each state 
are considered to conform to the following tendencies in most cases. 1)
From the transient states, one of all transitions with the arrows
directing to any other of the states is realized randomly. 2) From the
attractors, only a transition in the direction with the smallest 
$|Q(x \to y ; x_1, x_2)|$ ($= |Q_{min}|$) is realized with the
probability $\sim \exp(-Q_{min}/T)$ (escape rate). It is noted
that the system tends to stay the attractors for most of the time. Thus,
the escape rate from these attractors contributes dominantly to give the
mean velocity of the driven particle $u$. 

Now, we focus on the typical trajectories in the transition diagrams for 
the cases with $V < I/3$ (Figs. 4) to explain the appearance of the 
$F - u$ relations like in Fig. 2(a). When $F < V$, the transition
diagram as given in Fig. 4(a) involves the following trajectory as a
typical one characterizing the steady state motion, $(1,2) \to (1,3) \to (2,3) \to (2,0) \to (2,1) \to (3,1) \to (3,2) \to (0,2) \to (1,2)$.
(as indicated by thin arrows). 
This trajectory indicates the motions in which both particles proceed
in the direction of $F$. Here, this involves four 
attractors, $(1,2)$, $(2,3)$, $(2,1)$ and $(3,2)$. The escape rates from
the attractors $(1,2)$ or $(2,3)$ given through the transitions 
$(1,2) \to (1,3)$ or $(2,3) \to (2,0)$ increase with the increase in
$F$. However, those of $(2,1)$ or $(3,2)$ given through the transitions
$(2,1) \to (3,1)$ or $(3,2) \to (0,2)$ are independent of $F$. Then, $u$
cannot increase so fast with $F$ in this case. 

In the case where $V < F < I-2V$, the typical trajectories giving the 
steady state motions still involve the attractors $(2,1)$ and
$(3,2)$ as shown in Fig. 4(b) (as indicated by thin arrows). Also in
this case, the escape rates from these attractors are still independent
of $F$. In this case, $u$ increases little with $F$.

In the case where $I-2V < F < I-V$, the following two types of typical
trajectories appear, as shown in Fig. 4(c), i) trajectories including
two attractors, $(2,1)$ and $(3,2)$, for example 
$(1,2) \to (1,3) \to (2,3) \to (2,0) \to (2,1) \to (2,2) \to (3,2) \to (0,2) \to (1,2)$ (as indicated by thin arrows), 
and ii) trajectories including only one attractor $(2,1)$ for example 
$(2,1) \to (2,2) \to (2,3) \to (2,0) \to (2,1)$ (as indicated by broken
arrows). Here, the former trajectory indicates the motions in which both
particles proceed in the direction of $F$ while the latter indicates the
motions in which only the driven particle proceeds. In this case, the
escape rate from $(2,1)$ given through the transition $(2,1) \to (2,2)$
increases exponentially with the increase in $F$ while the escape rate
$(3,2)$ given through the transition $(3,2) \to (0,2)$ is independent of
$F$. Then, the mean velocity of the driven particle remains unaffected
by the increase in $F$ if the system develops along trajectories i). On
the other hand, this increases exponentially with $F$ if the system
develops along trajectories ii). Thus, within this range of $F$, $u$
increases steeply with $F$ on the average.  

In the case where $I-V < F < I$, similar to the previous case, the
following two types of trajectories are still typical ones as shown in
Fig. 4(d), i) trajectories including two states, $(2,1)$ and $(3,2)$ (as
indicated by thin arrows), and ii) trajectories including a state
$(2,1)$, but not including a state $(3,2)$ (as indicated by broken arrows).
However, the state $(2,1)$ is no longer the attractor although $(3,2)$
is still an attractor. Then, the diagram includes trajectories involving
an attractor $(3,2)$ and those involving no attractors. In the former
trajectories, the escape rate from $(3,2)$ is still independent of
$F$. Therefore, $u$ increases slightly with $F$. 

In the case where $I < F (< I+V)$, the escape rate from the attractor
$(3,2)$ becomes one increasing exponentially with $F$, which is
given through the transition $(3,2) \to (3,3)$ as in Fig. 4(e). Then, the
mean velocity of the driven particle along trajectories including
the attractor $(3,2)$ (as indicated by thin arrows) increases exponentially
with $F$, although that along any trajectories without the attractor
$(3,2)$ (as indicated by broken arrows) are unaffected by
$F$. Therefore, within this range of $F$, $u$ increases steeply with $F$. 

Thus, for $V < I/3$, $u$ increases slowly $\to$ steeply $\to$
slowly $\to$ steeply with the increase in $F$. This result is consistent
to the $F- u$ relations in Fig. 2(a).

In cases with $I/3 < V < I/2$ or $I/2 < V$, the manner of the $F$ 
dependent changes of the transition diagrams proceed differently 
from the previous case. 
(We do not show the diagrams for such $V$ in order to avoid the
complications.) 
We can also explain the mechanism of
the phenomena observed in each range of $V$ via the same analytic method.

\section{Summary and discussions}
In this paper, we investigated the transport phenomena of a simple 
nonequilibrium lattice gas system. We measure the mean velocity of the
driven particle ($u$) and the coefficient of effective drag of this
particle ($\eta$). 
Under the certain ranges of parameters, we obtained the $u$ - $\eta$
relations qualitatively similar to the relations which observed between
the shear rate and shear viscosity coefficient in the shear-thickening
polymer solutions\cite{poly0,poly1,poly11} or that of the shear-thickening
colloidal suspensions\cite{col0,col1}. The mechanisms of these phenomena
are explained by the transition diagram. 

Now, we consider the similarities between our results and those
obtained recent studies for the rheology of soft matter systems 
in a little more detail. The soft glassy systems like
polymer solutions or colloidal suspensions under shear have been studied
extensively by experiments, molecular dynamics simulations, the analysis
of the transient network model or the random spin model, and studies of
the mode coupling theory or an analysis of the pair distribution
function\cite{poly0,poly1,poly11,poly2,poly3,poly4,poly5,col0,col1,col2,col22,col3,col4}.

Among results of some experimental and theoretical studies for
shear-thickening polymer solutions, it has been observed that the shear 
viscosity varies in the form, increase $\to$ fast decrease $\to$
slow decrease $\to$ fast decrease, with the increase in the shear
rate\cite{poly0,poly3}. Such a profile was obtained in the $u - \eta$
relation in our simulation under a certain condition (for the case I-ii
in the phase diagram Fig. 3(d)). On the other hand, among results of
some experimental and theoretical studies for shear-thickening colloidal
suspensions, it has been observed that the shear viscosity varies in the 
form, decrease $\to$ increase $\to$ decrease, with the increase in the
shear rate\cite{col0,col1,col2,col22,col3,col4}. Such a profile was also
obtained in the $u - \eta$ relation in our simulation under the other
certain conditions (for the case II-ii or III in the phase diagram
Fig. 3(d)).

Moreover, among results of recent studies, the power law [shear viscosity
coefficient] $\sim$ [shear rate]$^{-\delta}$ has been observed in the
shear-thinning regime. Here, $\delta \sim 2/3$ has been given in some
experimental studies, simulations and theoretical
studies\cite{col1,col2,col22,col4} while $\delta$ that varies depending
on several conditions between $0$ and $1$ has been obtained in the other
experimental and theoretical studies\cite{poly2,col0,col3}. On the other
hand, we obtained the following $u - \eta$ relations in the $\eta$
decreasing regime for lower $T$; $\eta \sim u^{0 \sim -0.5}$ for smaller
$u$ and $\eta \sim u^{-0.9}$ for larger $u$. 

Thus, we expect our results to provide important hints to uncover 
the possible mechanism for several rheological characteristics of
several soft materials. 
However, the relations of some properties, for example the 
exponents of the transport coefficients, the transport properties in
cases with lower temperature, etc. are still unclear between our
results and those obtained in several recent studies.  
Then, studies of extended models,
including more lanes or particles in the space with more sites or
continuous space, and the comparisons between such toy systems and either
real systems or more realistic models of the polymer or colloidal
systems represent important future issues.

The author thanks to M. Sano, and M. Otsuki for useful 
discussions. This research was supported in part by a Grant-in-Aid for 
JSPS Fellows (10039).

\newpage

\begin{figure}
\begin{center}
%\psbox[width=8.0cm]{}
\includegraphics[width=8.0cm]{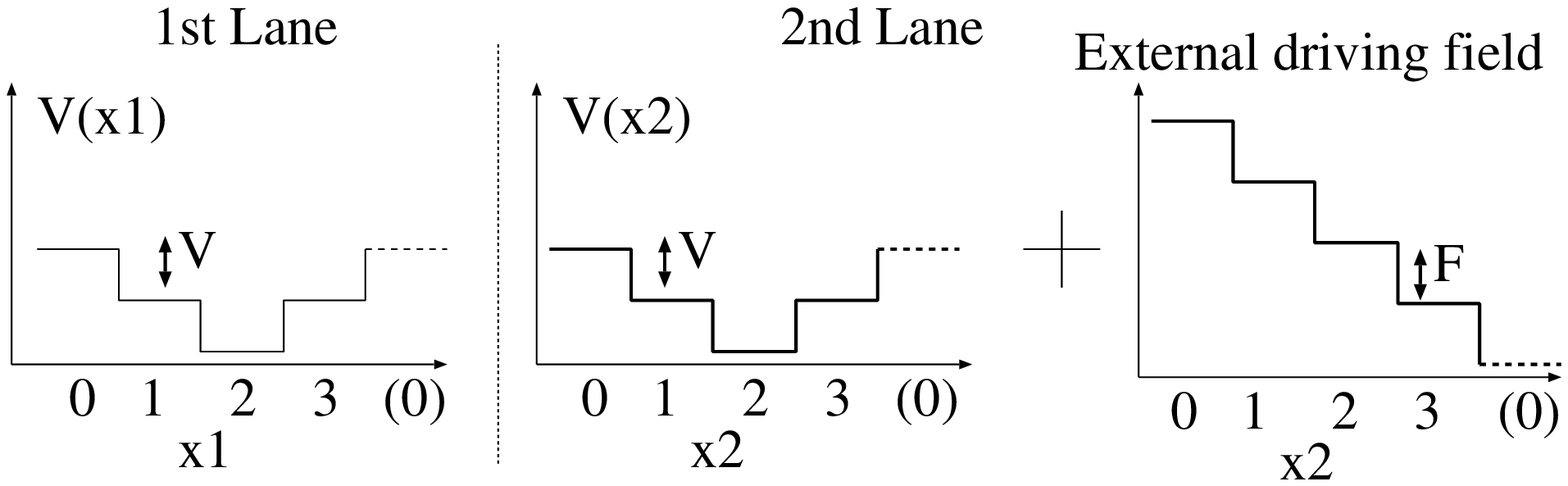}
\end{center}
\caption{Illustrations of effects of potential and external field in each lane.}
\end{figure}

\begin{figure}
\begin{center}
%\psbox[width=8.0cm]{}
\includegraphics[width=6.0cm]{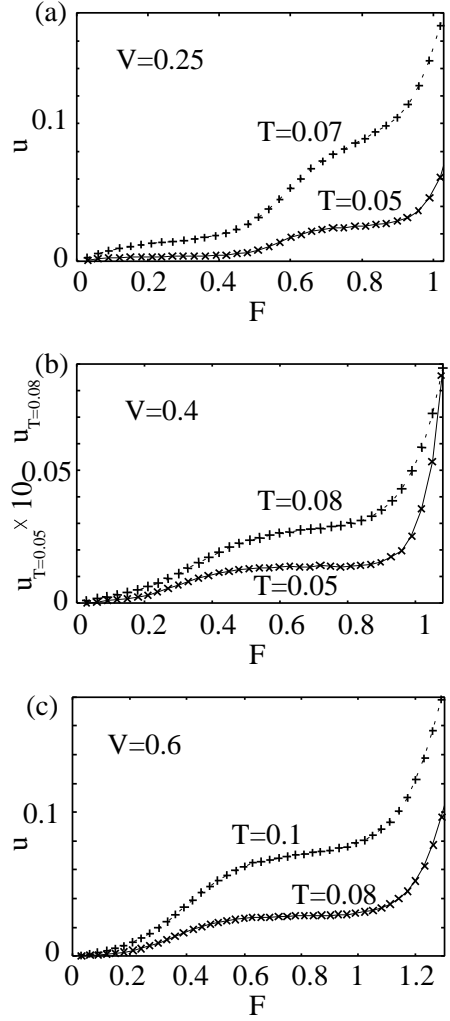}
\end{center}
 \caption{Mean velocity $u$ as a function of $F$ for 
(a) $V = 0.25$ with $T = 0.05$ or $T = 0.07$, 
(b) $V = 0.4$ with $T = 0.05$ or $T = 0.08$ and (c) $V = 0.6$ with 
$T = 0.08$ or $T = 0.1$.}
\end{figure}

\begin{figure}
\begin{center}
%\psbox[width=8.0cm]{}
\includegraphics[width=6.0cm]{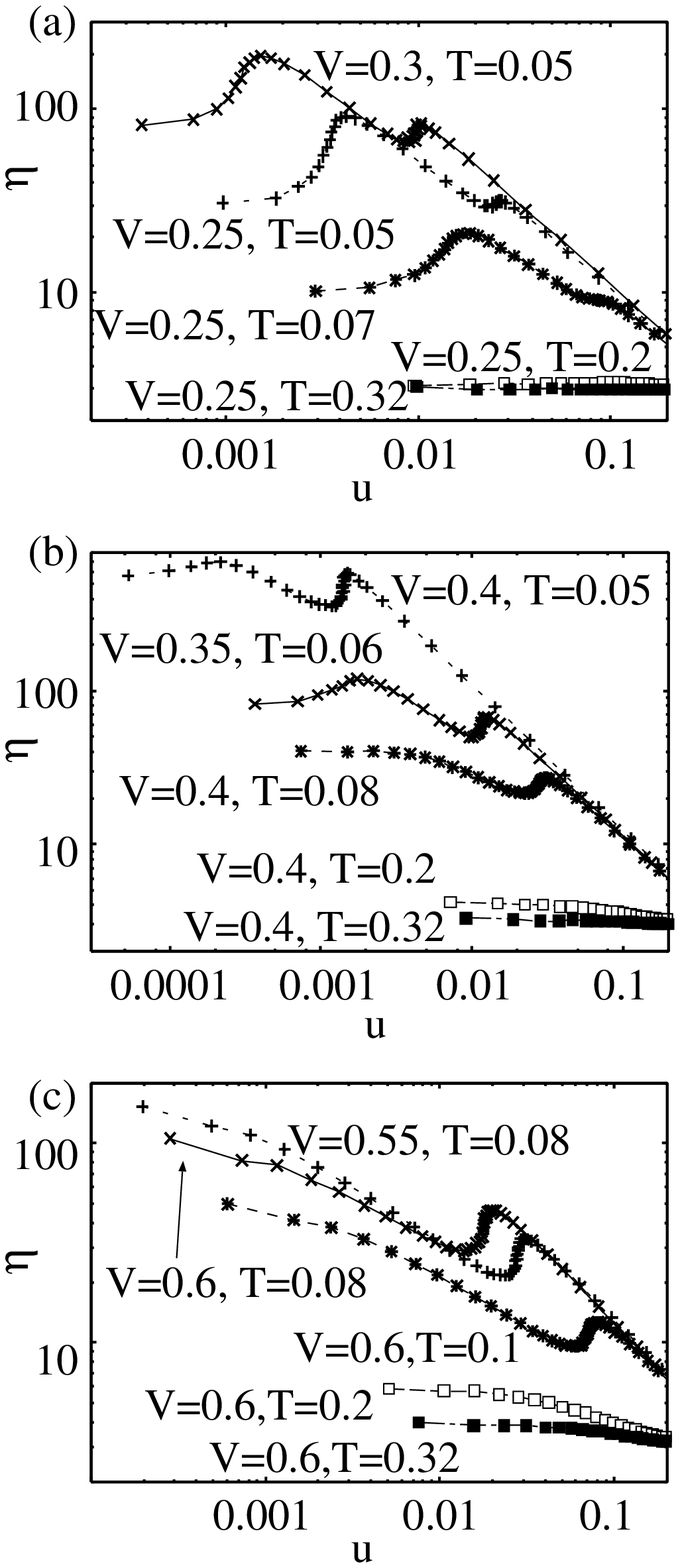}
\includegraphics[width=7.0cm]{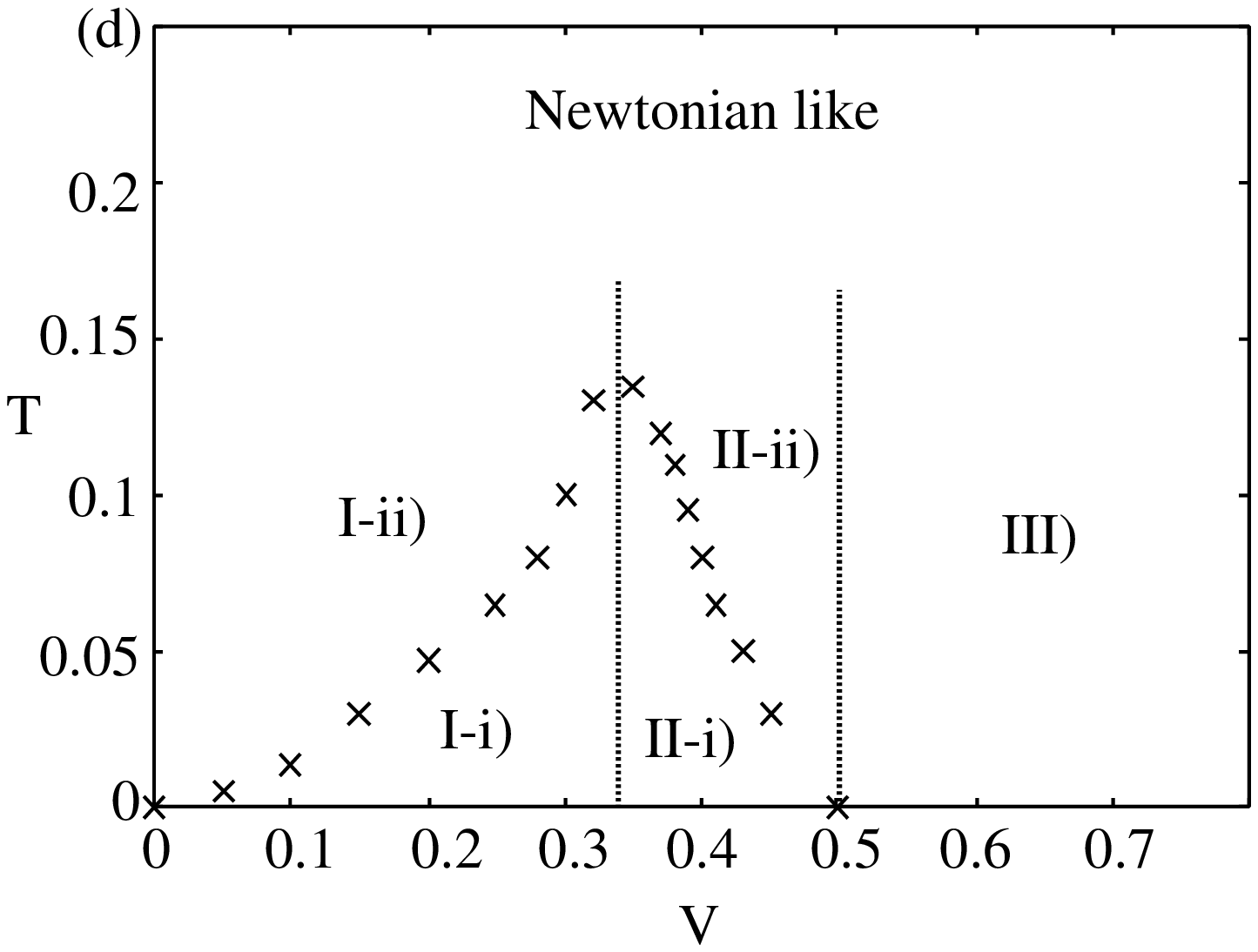}
\end{center}
 \caption{Coefficient of effective drag $\eta$ as a 
function of $u$ for (a) $V = 0.3$ with $T = 0.05$ (Times) and $V = 0.25$
 with $T = 0.05$ (Plus), $T = 0.07$ (Star), $T = 0.2$ (Empty square) or
 $T = 0.32$ (Full Square), (b) $V = 0.35$ with $T = 0.06$ (Times) and
 $V = 0.4$ with $T = 0.05$ (Plus), $T = 0.08$ (Star), $T = 0.2$ (Empty
 square) or $T = 0.32$ (Full Square) and (c) $V = 0.55$ with $T = 0.08$
 and $V = 0.6$ (Times) with $T = 0.08$ (Plus), $T = 0.1$ (Star), 
$T = 0.2$ (Empty square) or $T = 0.32$ (Full Square). (d) Phase diagram
 of the system as functions of $V$ and $T$. Dots and broken lines give
 the boundaries between states    
 }
\end{figure}

%\begin{center}
%%\psbox[width=8.0cm]{}
%\includegraphics[width=8.0cm]{AWA_E_FIG4a.ps}
%\includegraphics[width=8.0cm]{AWA_E_FIG4b.ps}
%\includegraphics[width=8.0cm]{AWA_E_FIG4c.ps}
%\includegraphics[width=8.0cm]{AWA_E_FIG4d.ps}
%\includegraphics[width=8.0cm]{AWA_E_FIG4e.ps}
%\end{center}

\begin{figure}
\begin{center}
\includegraphics[width=8.0cm]{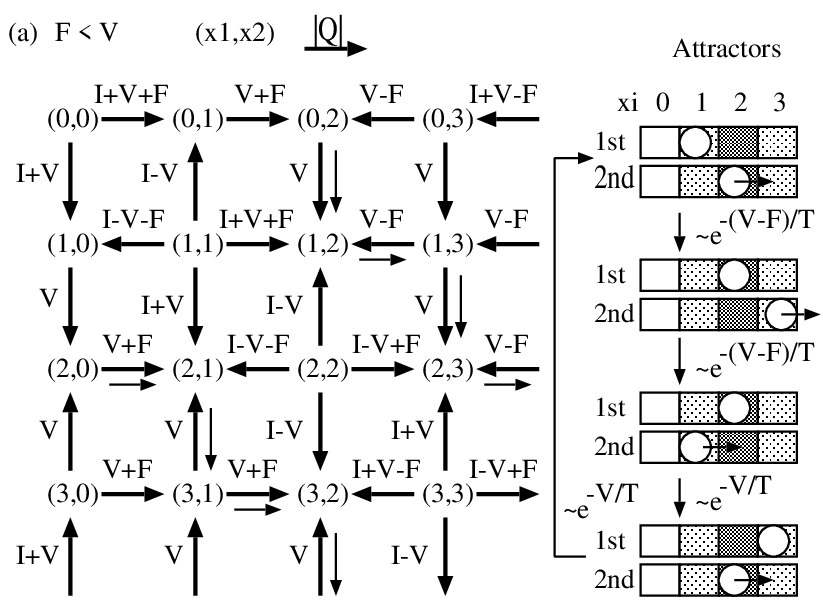}
\includegraphics[width=8.0cm]{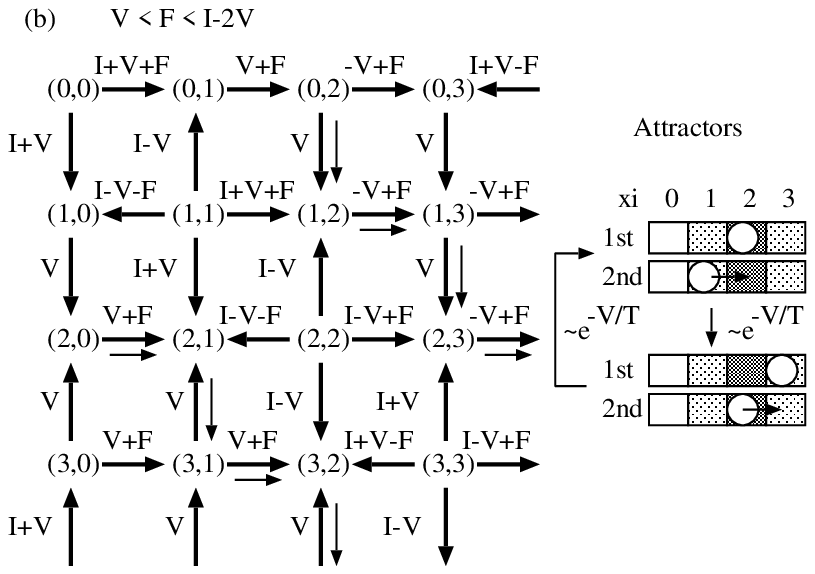}
\includegraphics[width=8.0cm]{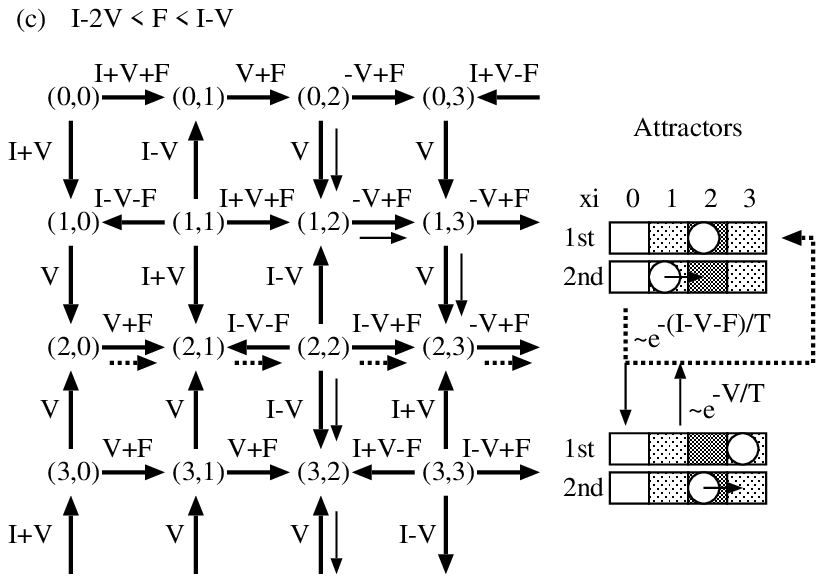}
\includegraphics[width=8.0cm]{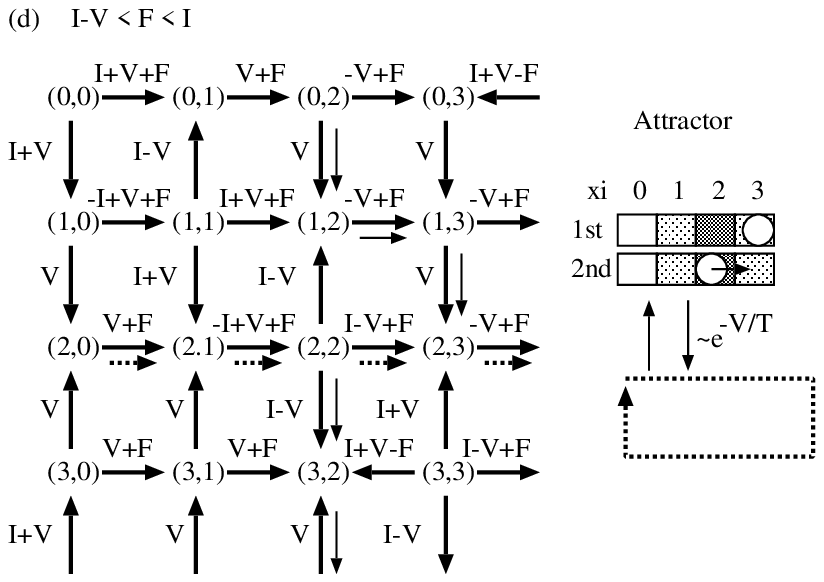}
\includegraphics[width=8.0cm]{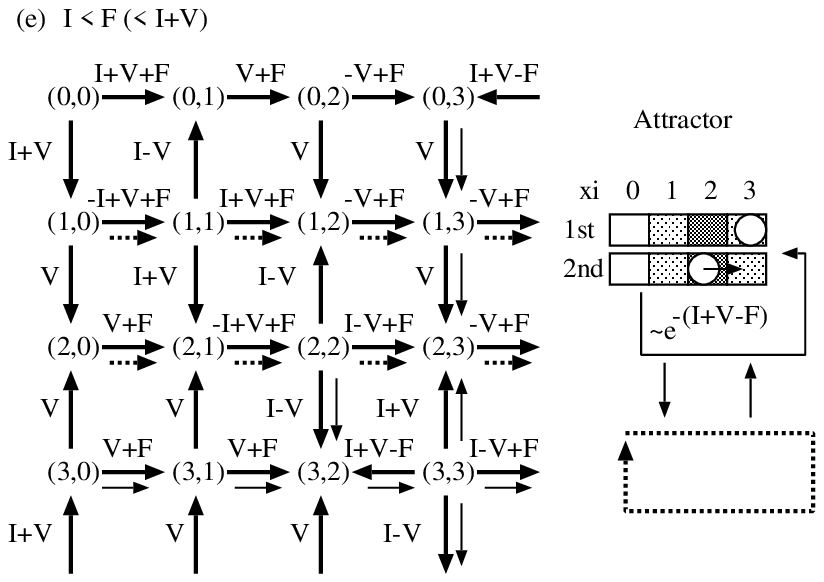}
\end{center}
\caption{Typical transition diagram (left), and illustrations of
 attractors and rough estimated escape rates from them (right) for 
(a)$F < V$, (b)$V < F < I-2V$, (c)$I-2V < F < I-V$, 
(d)$I-V < F < I$ and (e)$I < F < I+V$ under $V < I/3$.
}
\end{figure}

%\begin{center}
%%\psbox[width=8.0cm]{}
%\includegraphics[width=5.5cm]{AWA_E_FIG5a.ps}
%\includegraphics[width=5.5cm]{AWA_E_FIG5b.ps}
%\includegraphics[width=5.5cm]{AWA_E_FIG5c.ps}
%\includegraphics[width=5.5cm]{AWA_E_FIG5d.ps}
%\includegraphics[width=5.5cm]{AWA_E_FIG5e.ps}
%\end{center}
%
%\begin{figure}
%\begin{center}
%%\includegraphics[width=8.0cm]{AWA_E_FIG4a.ps}
%%\includegraphics[width=8.0cm]{AWA_E_FIG4b.ps}
%%\includegraphics[width=8.0cm]{AWA_E_FIG4c.ps}
%%\includegraphics[width=8.0cm]{AWA_E_FIG4d.ps}
%%\includegraphics[width=8.0cm]{AWA_E_FIG4e.ps}
%\end{center}
%\caption{Typical transition diagram for (a)$F < I-2V$, (b)$I-2V < F < V$
% and (c)$V < F < I-V$ under $I/3 < V < I/2$, and for $I/2 <
% V$ with (d)$F < V-I/2$ and (e)$V-I/2 < F < I-V$ under $I/2 < V$.
%}
%\end{figure}

\end{document}